\documentclass[twocolumn,superscriptaddress,aps,preprintnumbers,amsmath,amssymb,prl,footinbib]{revtex4-1}

\usepackage[titletoc,toc,title]{appendix}
\usepackage{titletoc}
\usepackage{titlesec}

\usepackage{textgreek}
\usepackage{amsmath}
\usepackage{amssymb}
\usepackage{amsfonts}
\usepackage{graphicx}
\usepackage{xcolor}
\usepackage{xfrac}
\usepackage{footmisc}
\usepackage{comment}
\usepackage{pifont}
\usepackage{times}
\usepackage[hidelinks]{hyperref}
\usepackage{,bm}
\usepackage{enumitem}
\usepackage{centernot}
\usepackage{cancel}
\usepackage{slashed}
\usepackage{relsize}
\usepackage{amsmath,amssymb,mathrsfs}
\usepackage{mdframed}
\usepackage{tcolorbox}
\usepackage{bm}
\usepackage{times}
\usepackage{epsfig}
\usepackage{verbatim}
\usepackage{bm}
\usepackage[utf8]{inputenc}
\usepackage{graphics}
\usepackage{graphicx,epsfig,amssymb,amsmath,color,cancel}
\usepackage{subfigure}
\usepackage[english]{babel}
\usepackage{adjustbox}
\usepackage{soul}
\usepackage[normalem]{ulem}
\usepackage{orcidlink}

\definecolor{zima_blue}{HTML}{1393C1}
\hypersetup{setpagesize=false,bookmarksnumbered=true,bookmarksopen=true,%
colorlinks=true,linkcolor=zima_blue,urlcolor=zima_blue,citecolor=zima_blue,linktocpage=false}

\usepackage{physics}

\newtcolorbox{mybox}[3][]{
  colframe = black,         
  colback  = #2!10,
  coltitle = #2!20!black,
  title    = {#3},
  boxsep   = 2pt,           
  top      = -6pt,           
  bottom   = 3pt,           
  #1,
}

\definecolor{rossoferrari}{HTML}{D9073D}
\definecolor{mediumblue}{HTML}{0000CD}
\definecolor{forestgreen}{HTML}{228B22}
\definecolor{desy_blue}{HTML}{009EE2}
\definecolor{desy_orange}{HTML}{FD8800}
\definecolor{peera_green}{HTML}{008B8B}
\definecolor{peera_orange}{HTML}{B22222}
\definecolor{light_pink}{rgb}{1,0.4,0.4}
\definecolor{light_blue}{rgb}{0.284602,0.317763,0.963947}
\definecolor{peera_col}{RGB}{240, 94, 28}
\definecolor{blue_col}{RGB}{0,92,175}
\definecolor{red_col}{RGB}{203,64,66}
\usepackage{hyperref}
\hypersetup{linktocpage,
    colorlinks=true,
    linkcolor=red_col,
    filecolor=Mahogany,      
    urlcolor=blue_col,
    citecolor=blue_col,
}

\def\MPl{M_\text{Pl}}

\begin{document}

\title{New Source for QCD Axion Dark Matter Production: Curvature Induced}

\author{Cem Er\"{o}ncel~\orcidlink{0000-0002-9308-1449}}
\email{cem.eroncel@istinye.edu.tr}
\affiliation{\.{I}stinye University, Faculty of Engineering and Natural Sciences, 34396, \.{I}stanbul, T\"{u}rkiye}
\author{Yann Gouttenoire~\orcidlink{0000-0003-2225-6704}}
\email{yann.gouttenoire@gmail.com}
\affiliation{School of Physics and Astronomy, Tel-Aviv University, Tel-Aviv 69978, Israel}
\affiliation{PRISMA+ Cluster of Excellence $\&$ MITP, Johannes Gutenberg University, 55099 Mainz, Germany}
\author{Ryosuke Sato~\orcidlink{0000-0003-2745-4208}}
\email{rsato@het.phys.sci.osaka-u.ac.jp}
\affiliation{Department of Physics, The University of Osaka, Toyonaka, Osaka 560-0043, Japan}
\author{G\'{e}raldine Servant~\orcidlink{0000-0002-3063-363X}}
\email{geraldine.servant@desy.de}
\affiliation{Deutsches Elektronen-Synchrotron DESY, Notkestra{\ss}e 85, 22607 Hamburg, Germany}
\affiliation{II. Institute of Theoretical Physics, Universit\"{a}t Hamburg, 22761, Hamburg, Germany}
\author{Peera Simakachorn~\orcidlink{0000-0002-4274-1179}}
\email{peera.sima@gmail.com}
\affiliation{IFIC, Universitat de Val\`{e}ncia-CSIC,
C/ Catedr\`{a}tico Jos\'{e} Beltr\`{a}n 2, E-46980, Paterna, Spain}
\affiliation{Khon Kaen Particle Physics and Cosmology Theory Group (KKPaCT),
Department of Physics, Faculty of Science, Khon Kaen University,
123 Mitraphap Rd., Khon Kaen, 40002, Thailand}

\date{\today}

\preprint{DESY-25-033}
\preprint{MITP-25-018}
\preprint{OU-HET-1265}

\date{\today}


\begin{abstract}
We discuss a novel mechanism for generating dark matter from a fast-rolling scalar field, relevant for both inflation and rotating axion models, and apply it specifically to the (QCD) axion. Dark matter comes from scalar field fluctuations generated by the product of the curvature perturbation and the fast-rolling background field. These fluctuations can explain the totality of dark matter in a vast axion parameter space, particularly for the QCD axion, which will be targeted by upcoming experiments. We review the constraints on this mechanism and potential gravitational-wave signatures.

\vspace{0.3cm}
\noindent
\textit{Published version:~\href{https://doi.org/10.1103/PhysRevLett.135.231002}{Phys. Rev. Lett. 135, 231002}}
\end{abstract}

\maketitle

\textit{Introduction.}--Axion-like particles (ALP) are leading candidates to explain the dark matter (DM) of the Universe \cite{Preskill:1982cy,Abbott:1982af,Dine:1982ah}.
Several mechanisms to produce ALPs have been advocated: thermal production in the case of large axion couplings to the Standard Model as is the case for the QCD axion with a low decay constant $f_a$ \cite{DEramo:2022nvb}, standard misalignment \cite{Preskill:1982cy,Abbott:1982af,Dine:1982ah}, large misalignment \cite{Arvanitaki:2019rax},  kinetic misalignment \cite{Co:2019jts,Chang:2019tvx,Co:2020dya,Co:2020jtv,Co:2020xlh}. In the latter case, it was shown that in most of the parameter space, the axion field is fragmented and the axion energy density is not confined in the zero-mode \cite{Fonseca:2019ypl,Eroncel:2022vjg,Eroncel:2022efc,Eroncel:2024rpe}. Axions can also be produced from the emission of the cosmic string-domain wall network \cite{Gorghetto:2018myk,Gorghetto:2020qws,Gorghetto:2022ikz,Gorghetto:2023vqu}.
Axions from standard misalignment and from cosmic strings are the most studied cases, although they correspond to a region of parameter space in the [axion mass $m_a$, axion coupling $\propto f_a^{-1}$]  plane that is difficult to probe experimentally. In contrast, kinetic misalignment followed by axion fragmentation, as arising in models of rotating axions,  opens a large region of parameter space for ALP DM that will be tested in upcoming experiments.

This \textit{letter} presents a completely novel mechanism for scalar particle production in the early Universe, that can be applied generally to scalar DM, and specifically to rotating axion models. Rather than coming from the energy initially stored in the zero-mode background field, the energy density comes from the fluctuations of the scalar field that are induced by primordial (inflationary) curvature perturbations, see also~\cite{Maleknejad:2024ybn,Maleknejad:2024hoz,Garani:2024isu,Garani:2025qnm}. Such a source term is effective if the background scalar field has a large kinetic energy.

\begin{figure}[t!]
\centering
\includegraphics[width=0.48\textwidth]{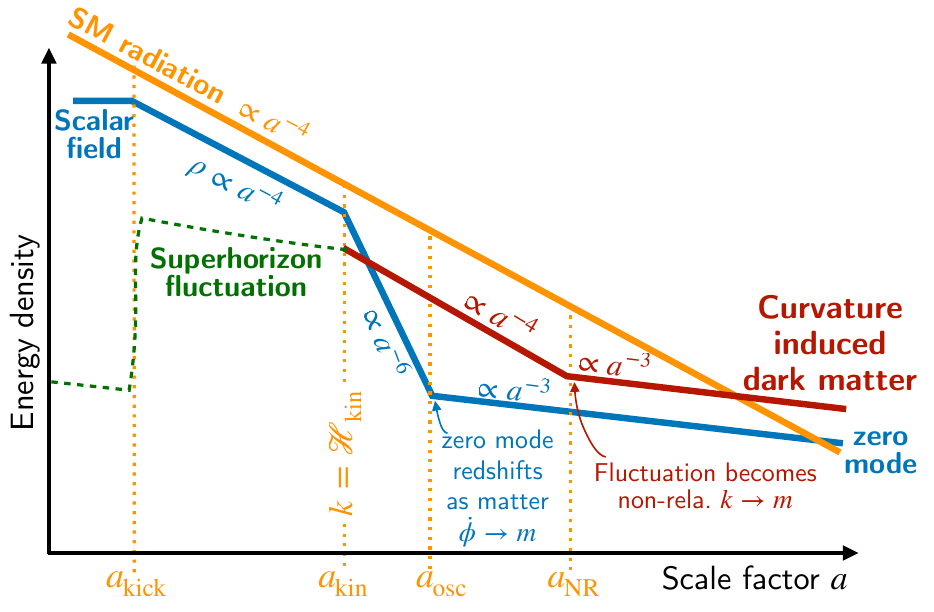}
\caption{
Cartoon of the energy-density evolution of a fast-moving scalar field condensate in blue, which behaves as radiation $\sim a^{-4}$, kination $\sim a^{-6}$, and matter $\sim a^{-3}$, consecutively, during the radiation era of the Universe. The second stage starts at $a_{\rm kin}$ when the field dynamics is kinetic-energy dominated. The last stage happens when the field gets trapped and oscillates inside the potential barrier at $a_{\rm osc}$, leading to DM via kinetic misalignment or fragmentation. The red line shows the energy-density evolution of scalar-field fluctuations $\rho_{\rm fluct}$ induced by primordial curvature perturbations. Initially produced relativistically, they become non-relativistic at $a_{\rm NR}$,  and can serve as DM.
The orange line shows the Standard Model radiation.
Another possible scenario is if the scalar field is the inflaton itself and fast-rolls after inflation, that would correspond to setting $a_{\rm kick}=a_{\rm kin}$. 
Plot is schematic and not to scale.
}
\label{fig:cartoon}
\end{figure}

The generation of these fluctuations was pointed out in ~\cite{Eroncel:2022vjg,Apers:2024ffe}, and some important implications, when applied to massless fields, were shown in \cite{Eroncel:2025bcb}. 
In this work, we study how they can behave as DM, as suggested in~\cite{Eroncel:2024rpe}. Depending on the scalar mass value as well as other UV parameters, their contribution may be the dominant one over the abundance from the zero mode, as shown in Fig.~\ref{fig:cartoon}. 
This general mechanism could have other applications. For instance, if the scalar field is the inflaton itself and inflation is ended by a fast-rolling stage (like in quintessential inflation \cite{Peebles:1998qn,Dimopoulos:2001ix} and string cosmologies \cite{Apers:2024ffe}),  fluctuations can be efficiently produced and could explain DM. 
The next two sections are rather general and model-independent, while the last section applies to a model of rotating axions.

\begin{figure}[t!]
    \centering
    \includegraphics[width=\linewidth]{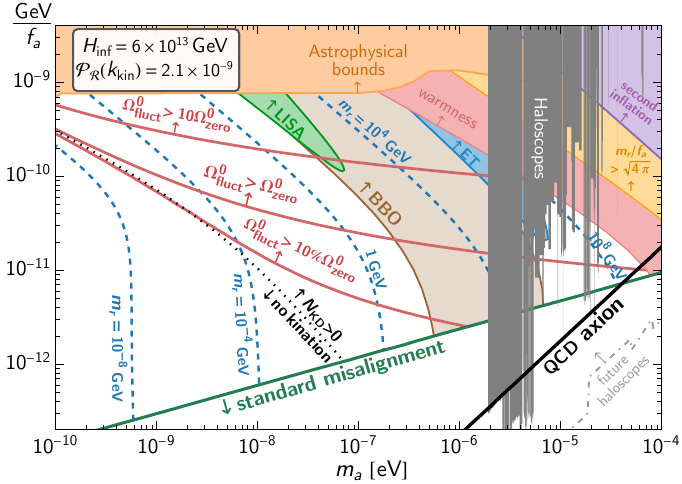}\\[-1em]
    \caption{
     Detectability of the kination peak of inflationary gravitational waves at LISA (green), BBO (brown), and ET (blue), using the scalar potential (\ref{quadraticpotential}) with $\{\kappa,l\} = \{10^{-4},13\}$.  
      Yellow region violates perturbativity  ($m_a/f_a>\sqrt{4\pi}$). Purple region is excluded by the second inflation bound (Eq.(8.22) of \cite{Gouttenoire:2021jhk}). Pink region has DM too warm [$\omega_{\rm fluct}^{\rm eq}\simeq k_{\rm kin}/(3a_{\rm eq}m_a)>10^{-8}$]. Red lines represent the ratio $\Omega_{\rm fluct}^0/\Omega_{\rm zero}^0$. Below the black dotted line, there is no kination era; the curvature-induced DM is produced in radiation era and becomes less abundant due to $\Omega_\phi(\eta_{\rm kin})<1$ in Eq.~\eqref{eq:yield_ratio}.        }
    \label{fig:kination_domination}
\end{figure}

\textit{Curvature-induced fluctuations of fast-rolling field.}--In the absence of anisotropic stress, the space-time metric in the conformal Newtonian gauge is 
\begin{equation}
\label{eq:metric}
ds^2=a^2\Big[-(1+2\Phi)d\eta^2+(1-2\Phi)\delta_{ij}dx^idx^j\Big],
\end{equation}
where $\Phi$ is the scalar metric fluctuation whose Fourier modes are defined by $\Phi_{\mathbf{k}}=\int d^3\mathbf{x}\,e^{i\mathbf{k}\cdot\mathbf{x}}\Phi({\mathbf{x}})$, and $\eta$ is conformal time. On super-horizon modes, $\Phi$ is frozen and related to the gauge-invariant comoving curvature perturbation  $\mathcal{R}\equiv -\Phi - {2}{(\mathcal{H}^{-1}\Phi'+\Phi)}/{3(1+\omega)} \simeq -3\Phi/2$, where $(\cdot)' \equiv d(\cdot)/d\eta$, $\mathcal{H}=a H$ is the comoving Hubble rate and $\omega$ is the Universe equation of state. We assume radiation-domination $\omega=1/3$ (unlike \cite{Eroncel:2025bcb}) and treat the case of kination-domination in Fig.~\ref{fig:kination_domination}.

Consider a fast‐rolling scalar field, $\phi=\overline{\phi}+\delta\phi$, with $\overline{\phi}$ its homogeneous part and $\delta\phi$ its fluctuations, as already discussed in \cite{Eroncel:2025bcb}.
During fast-rolling -- when kinetic and gradient energies dominate and potential energy is negligible -- and under the conditions $\Phi,\delta\phi\ll 1$, the scalar field equation of motion becomes
\begin{equation}
\delta\phi_{\mathbf{k}}''+2\mathcal{H}\,\delta\phi_{\mathbf{k}}'+k^2\delta\phi_{\mathbf{k}}= 4\Phi_{\mathbf{k}}'\overline{\phi}'.
\end{equation}
The right-hand side of this equation is the key point. The non-zero scalar speed $\overline{\phi}'$ combined with the curvature perturbation $\Phi_{\mathbf{k}}'$ sources the scalar fluctuation with momentum $k$, leading to the sub-horizon solution ($k\gg\mathcal{H})$
\begin{equation}
\label{eq:EOM_sol}
\phi_{\mathbf{k}}(\eta)\simeq \alpha\,\mathcal{R}_{\mathbf{k}}(0)\frac{\overline{\phi}'}{\mathcal{H}}\cos(k\eta),
\end{equation}
for modes entering during the fast-rolling stage $k<\mathcal{H}_{\rm kin}$ and $\alpha \simeq 0.52$ (see \cite[App.~B.2]{Eroncel:2022vjg} and App.~\ref{app:derivation_rho_fluct} for a rederivation).
At all orders in $\delta \phi$, the scalar field energy density is $\rho_\phi=[\left({\phi'}\right)^{\!2}+\left({\nabla\phi}\right)^{\!2}]/(2a^2)=\overline{\rho}_\phi(1+\delta_\phi+{\delta_\phi^2}/{2})$,
with $\overline{\rho}_\phi=(\overline{\phi}'/a)^2/2$ and the linear component of the density contrast $\delta_\phi\equiv (\rho_{\phi}/\overline{\rho}_\phi-1)_{\rm linear}=2\delta\phi'/\overline{\phi}$.
We deduce the volume-average of the energy density fluctuations~\cite{Eroncel:2025bcb}
\begin{equation}
\label{eq:rho_fluct_1}
  \rho_{\rm fluct}(\eta)\equiv \left\langle  \rho_{\phi}-\overline{\rho}_\phi\right\rangle = \frac{\overline{\rho}_\phi(\eta)}{2}\int d\log{k}~ \mathcal{P}_{\delta_\phi}(k,\eta),
\end{equation}
where $\mathcal{P}_{\delta_\phi}(k)\equiv \int \!\!d^3\mathbf{x} \,  e^{i\mathbf{k}\!\cdot\mathbf{x}}\!\left< \delta_\phi(0)\delta_\phi(\mathbf{x})\right>$ is the dimensionless power spectrum of the linear density contrast. We used the important fact that $\left<\delta_\phi\right>\propto \left<\mathcal{R}_\mathbf{k} \right> =0$ for adiabatic primordial curvature perturbations. Using Eq.~\eqref{eq:EOM_sol} and averaging over oscillations, we get $\mathcal{P}_{\delta_\phi}\simeq 2\alpha^2\mathcal{P}_{\mathcal{R}}(k)\qty(k/\mathcal{H})^2$ and Eq.~\eqref{eq:rho_fluct_1} leads to the fractional energy density $\Omega_{\rm fluct}\equiv \rho_{\rm fluct}/\rho_{\rm tot}$ of the fluctuations of the fast-rolling field 
\begin{equation}
 \label{eq:rho_fluct_end_1}
     \frac{d\Omega_{\rm fluct}(\eta)}{d\log{k}} = \alpha^2\Omega_{\phi}(\eta_{\rm kin})  \left(\frac{k}{\mathcal{H}_{\rm kin}}\right)^{\!2}\!\mathcal{P}_{\mathcal{R}}(k),
\end{equation}
where $\mathcal{H}_{\rm kin}= a_{\rm kin} H_{\rm kin}$ is the value of the Hubble rate when the scalar field starts rolling at $a_{\rm kin}$.
We used $\mathcal{H}\propto a^{-1}$ and introduced $\Omega_{\phi}(\eta)\equiv \overline{\rho_{\phi}}/\rho_{\rm tot}$ which redshifts like $\propto a^{-2}$  during radiation domination. We now study the novel implications for DM.

\textit{Dark matter from curvature-induced scalar fluctuations.}--We consider a scalar field $\phi$ with mass $m$. Below a scale factor
$a_{\rm NR}(k)=k/m$, the mode $k$ turns non-relativistic and its energy density redshifts like matter. From integrating Eq.~\eqref{eq:rho_fluct_end_1} over all modes, we deduce the total fractional energy abundance of scalar fluctuation today 
\begin{equation}
 \label{eq:rho_fluct_end_2}
    \frac{\Omega_{\rm fluct}^0}{\Omega_{\rm rad}^{0}} = \alpha^2 \Omega_{\phi}(\eta_{\rm kin})\left(\frac{a_0 m}{\mathcal{H}_{\rm kin}}\right) \!\int_{k_{\rm osc}}^{k_{\rm kin}}\frac{dk}{k}\left(\frac{k}{\mathcal{H}_{\rm kin}}\right)\mathcal{P}_{\mathcal{R}}(k).
\end{equation}
where $h^2\Omega_{\rm rad}^{0} \simeq 4.21 \times 10^{-5}$~\cite{ParticleDataGroup:2024cfk} with $h\simeq 0.68$ the reduced Hubble factor and $a_0$ is today's scale factor. 
The expression at any time is provided in Eq.~\eqref{eq:rho-fluct-integral} of Supplemental Material \cite{SupplementalMaterial}.
We model the curvature power spectrum as a power law
\begin{equation}
\label{eq:Delta_R_k}
  \mathcal{P}_{\mathcal{R}}(k) = \mathcal{P}_{\mathcal{R}}(k_{\star})\left({k}/{k_{\star}}\right)^{n_s(k)-1},
\end{equation}
where the amplitude and spectral index are measured by Planck at the pivot scale $k_{\star}\equiv 0.05~\rm Mpc^{-1}$~\cite{Planck:2018vyg},
 $  \mathcal{P}_{\mathcal{R}}(k_{\star})\simeq 2.099(29)\times 10^{-9}, ~ n_s(k_{\star})\simeq 0.965(4)$.
For $k$ away from $k_{\star}$, we use the Taylor-expansion $n_s(k) = n_s(k_{\star})-(3.65 \times 10^{-4})\ln(k/k_{\star})$  obtained after analyzing 300 single-field inflation models \cite{Martin:2024nlo}.
Plugging Eq.~\eqref{eq:Delta_R_k} into Eq.~\eqref{eq:rho_fluct_end_2}, the total abundance for $\mathcal{H}_{\rm kin}\gg k_{\rm osc}$ is dominated by the mode $k_{\rm kin}$ that re-enters at $a_{\rm kin}$
\begin{equation}
 \label{eq:rho_fluct_end_final}
     \frac{\Omega_{\rm fluct}^0}{\Omega^0_{\rm rad}} =  \alpha^2\Omega_{\phi}(\eta_{\rm kin})\left(\frac{a_0 m}{\mathcal{H}_{\rm kin}}\right)
\frac{\mathcal{P}_{\mathcal{R}}(k_{\rm kin})}{n_s},
 \end{equation}
Note that ${\mathcal{P}_{\mathcal{R}}(k_{\rm kin})}$ is poorly constrained and could be in principle many orders of magnitude larger than ${\mathcal{P}_{\mathcal{R}}(k_{\star})}$, a fact exploited in inflationary models that predict primordial black holes \cite{Byrnes:2025tji}.
The abundance of fluctuations $\delta \phi_\mathbf{k}$ should be compared with the one of the zero mode $\overline{\phi}$
\begin{equation}
    \label{eq:rho_zero_end_final}
     \Omega_{\rm zero}^0 \simeq \Omega_{\rm rad}^{0} \Omega_{\phi}(\eta_{\rm kin})\left( \frac{a_0}{a_{\rm osc}}\right)\left(\frac{a_{\rm kin}}{a_{\rm osc}}\right)^2,
\end{equation}
where $a_{\rm osc}$ is the scale factor when the homogeneous scalar field stops fast-rolling and starts oscillating, redshifting like matter \cite{Co:2019jts,Chang:2019tvx}. 
This occurs when the scalar field energy density drops below $\overline{\rho_{\phi}}(\eta_{\rm osc})$. Denoting $\overline{\rho_{\phi}}(\eta_{\rm osc}) \simeq m_a^2f_a^2$ in view of applying this to axions, with mass $m$ denoted by $m_a$, and decay constant $f_a$, this leads to $a_{\rm kin}/a_{\rm osc}\simeq (m_a f_a)^{1/3}/\overline{\rho}^{1/6}_\phi(\eta_{\rm kin})$. We deduce 
\begin{equation}
    \frac{\Omega_{\rm fluct}^0}{\Omega_{\rm zero}^0} \simeq 0.9\,\frac{\alpha^2}{n_s}\Omega_{\phi}^{1/2}(\eta_{\rm kin}) \left(\frac{10^{10}\,{\rm GeV}}{f_a}\right)\left(\frac{\mathcal{P}_{\mathcal{R}}(k_{\rm kin})}{2.1\cdot 10^{-9}}\right).
    \label{eq:yield_ratio}
\end{equation}
The fluctuation energy density can be further enhanced by up to one order of magnitude when the scalar field dominates the Universe, inducing a kination era, see App.~\ref{app:yields_dominating_case} for a derivation. Such a situation can be natural in rotating axion models and is associated with large backgrounds of gravitational waves (GW) \cite{Gouttenoire:2021wzu,Gouttenoire:2021jhk,Co:2021lkc}. We illustrate the observability prospects at future GW observatories in  Fig.~\ref{fig:kination_domination}.

\begin{figure}[t!]
\centering
\includegraphics[width=\linewidth]{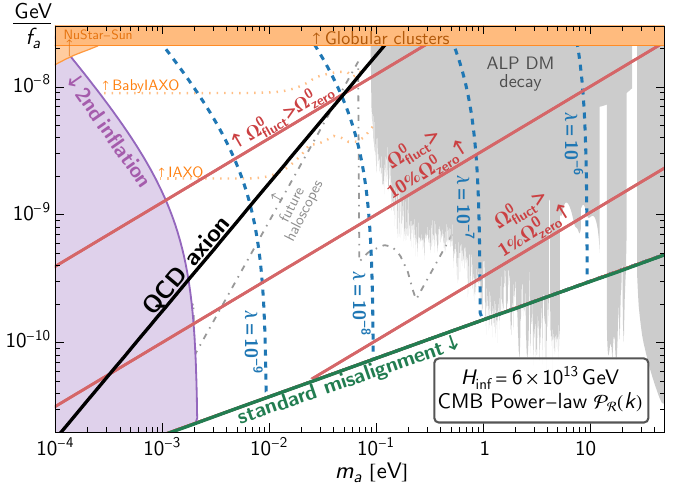}
\includegraphics[width=\linewidth]{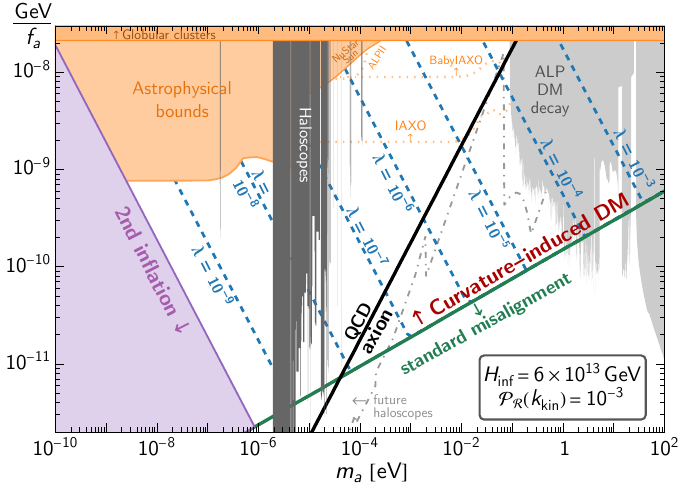}
\caption{
The rotating axion with quartic potential~\eqref{eq:complex_scalar_potential} produces the observed DM abundance in the region above the green line, via curvature-induced production and kinetic misalignment, i.e., $\Omega_{\rm DM}^0 = \Omega_{\rm fluct}^0 + \Omega_{\rm zero}^0$.
For each $\lambda$, DM is overproduced on the right of each dashed blue line.
The top panel assumes ${\mathcal{P}_{\mathcal{R}}(k)}$ using Eq.~\eqref{eq:Delta_R_k}. Red lines show the fraction of curvature-induced DM to zero-mode DM. The bottom panel assumes  $\mathcal{P}_{\mathcal{R}}(k_{\rm kin}) = 10^{-3}$. 
The purple region is ruled out by the second inflation bound \eqref{eq:second_inflation_bound}, while the warmness constraint \eqref{eq:warmness_bound} does not appear.
The gray regions are the combined bounds for ALP DM from haloscopes and ALP-decay searches \cite{Wang:2023imi,Todarello:2023hdk,Nakayama:2022jza,Carenza:2023qxh,Todarello:2024qci,Janish:2023kvi,Pinetti:2025owq,Saha:2025any,Liu:2023nct,Capozzi:2023xie,Wadekar:2021qae}, while the orange constraints apply to any ALP. The bounds from ALP-decay would become weaker if axion does not form DM \cite{Langhoff:2022bij}. Regions above gray and orange dotted and dot-dashed will be probed by future experiments. All constraints are compiled from \cite{AxionLimits}, assuming $g_{\phi \gamma \gamma} \simeq 1.02\alpha_{\rm em}/(2\pi f_a)$ as motivated by the KSVZ model \cite{Kim:1979if,Shifman:1979if}.
}
\label{fig:uv_complete}
\end{figure}

Fluctuations are produced with $ k \gg m $ and initially behave as radiation, so they are constrained by structure formation. The Lyman-$\alpha$ limit on thermal DM, $m_{\rm WDM}\gtrsim 2~\,\mathrm{keV}$~\cite{Garzilli:2019qki}, translates into an upper bound on the DM equation of state at matter–radiation equality, $\omega_{\rm DM, eq}\lesssim10^{-8}$~\cite{Ballesteros:2020adh}. In the non-relativistic limit ($ k\ll m $), the EoS for a mode is $\omega_{\rm fluct}\simeq k/(3\,a\,m)$, see App.~\ref{app:equation_of_state} for details. The dominant contribution arises from $ k\simeq k_{\rm kin} $, which leads to the bound
\begin{equation}
T_{\rm kin} \lesssim 1.1\times10^{15}\,\mathrm{GeV} \left(\frac{m}{1\,\mathrm{eV}}\right)\frac{g_{*s}^{1/3}(T_{\rm kin})}{g_*^{1/2}(T_{\rm kin})},
\label{eq:warmness_bound}
\end{equation}
where $T_{\rm kin}$ is the Universe temperature at $a_{\rm kin}$.
This condition is satisfied in the model parameter space of Fig.~\ref{fig:uv_complete}, but is modified if the scalar field dominates the Universe and leads to a constraint in Fig.~\ref{fig:kination_domination}.

Eq.~\eqref{eq:rho_fluct_end_1} suggests that the energy density fraction in radiation-like fluctuations is fixed over time and is proportional to $\Omega_{\rm \phi} \mathcal{P}_{\mathcal{R}}(k_{\rm kin}) \ll 1$. Therefore, the curvature-induced axion DM is exempted from the dark radiation bound imposed by Big-Bang Nucleosynthesis~\cite{Pitrou:2018cgg} and CMB~\cite{Planck:2018vyg}.

In the above analysis, we neglected the interactions between the curvature-and-fast-roll-induced fluctuations and the zero mode. 
We provide an estimate of this effect in Appendix~\ref{App:backreaction} for the rotating-axion model, which indicates that the zero-mode contribution could be further suppressed and the dominant relic would come from the fluctuations. 
Nonetheless, a dedicated numerical simulation would be needed for estimating the size of such corrections.

\textit{Application: DM from rotating axions.}--
We now apply our findings to a well-motivated framework realising fast-moving axions:
rotating complex fields, as discussed extensively in Ref.~\cite{Co:2019wyp,Co:2019jts,Co:2020dya,Co:2020jtv,Co:2021lkc,Gouttenoire:2021wzu,Gouttenoire:2021jhk,Eroncel:2022vjg,Eroncel:2024rpe}.
They rely on a complex scalar field $\chi = S e^{i\phi/f_a}$, corresponding to the Peccei-Quinn field in the case of the QCD axion \cite{Peccei:1977hh,Peccei:1977ur}, where $S$ and $\phi$ describe the radial and angular field directions.  The latter is identified with the axion. The key evolution is that the dynamics start at a large-$S$ value rather than at the origin of the potential. Such scenario requires four ingredients: \emph{i)} a scalar potential invariant under a global $U(1)$ symmetry with the spontaneous breaking minimum at $S= f_a$ that allows a motion along the axion direction at late times, \emph{ii)} an explicit $U(1)$--breaking term that operates only at early times and kicks the complex field into an elliptic orbit,
\emph{iii)} a large radial initial field value $S_{\rm ini}$ that enables the angular kick, \emph{iv)} a radial-motion damping mechanism to bring the field orbit into a circle, which later settles down to the potential minimum and starts the kination behavior (i.e., $\rho_\phi \propto a^{-6}$).
Here, we remain agnostic on the origins of some ingredients as they depend on the complete realization of such models, which were discussed extensively in e.g.~\cite{Gouttenoire:2021jhk,Eroncel:2024rpe}.

Let us consider for ingredient \emph{i)} a $U(1)$-invariant quartic potential and for ingredient \emph{iii)} a Hubble-induced mass term: 
\begin{equation}
    V(\chi) = \lambda(|\chi|^2 - f_a^2)^2 - \begin{cases}
    \lambda_H H^2 |\chi|^2 ~ &\textrm{during inflation}\\
    0 ~ ~ &\textrm{after inflation},
    \end{cases}
    \label{eq:complex_scalar_potential}
\end{equation}
which also suppresses the axion isocurvature perturbation.

During inflation, the field is driven to the potential minimum 
  $  S_{\rm ini} = f_a \sqrt{1 + \beta^2} ~ ~ \textrm{with} ~ ~ \beta^2 \equiv \lambda_H {H_{\rm inf}^2}/{(2\lambda f_a^2)},$
where $H_{\rm inf}$ is the Hubble scale during inflation. 
After inflation, we assume for simplicity that the Universe reheated instantaneously \footnote{In the case of non-instantaneous reheating with a period after inflation where the total energy density of the Universe redshifts as $\rho_{\rm tot} \sim a^{-n}$ --- with $n=3,4,6$ for matter, radiation, and kination eras --- when the Hubble parameter is $H_{\rm inf}> H > H_{\rm reh}$ and $H_{\rm osc} > H_{\rm reh}$, we find that $\bar{\rho}_{\phi,\rm kin}$ remains unchanged, as it depends on the dynamics of the scalar field. On the other hand, instead of Eq.~\eqref{eq:model_background_KD}, we have 
        $\rho_{\rm tot,kin}  
        = \rho_{\rm kick}^{\rm tot} \left({a_{\rm kick}}/{a_{\rm kin}}\right)^4 ({H_{\rm reh}}/{H_{\rm kick}})^{{2(n-4)}/{n}}$,
    where we used $H \propto a^{-n/2}$ for $H_{\rm inf}> H > H_{\rm reh}$. In this case, 
    we find that the final ratio of the DM abundances depends on the reheating scale $H_{\rm reh}$ as, 
    ${\Omega_{\rm fluct}^0}/{\Omega_{\rm zero}^0} \propto \left[{\sqrt{2}}{H_{\rm inf}}/{(3H_{\rm reh})}\right]^{1-4/n}$ where the extra factor will modify $\Omega_{\phi}^{1/2}(\eta_{\rm kin})$ in Eq.~\eqref{eq:yield_ratio}.
    The abundance of DM from axion fluctuation is suppressed when $n<4$, in particular for matter domination after inflation. On the contrary, the axion abundance is enhanced for a stiff era ($n>4$).}.
The scalar field energy density is set by the potential,
    $\rho_{\chi,\rm ini} = V(S_{\rm ini}) = \lambda f_a^4 \beta^4 = {\lambda_H^2 H_{\rm inf}^4}/{(4 \lambda)},$
and is subdominant with respect to the total radiation energy density of the Universe. 
The field remains frozen due to Hubble friction.

The complex field starts moving when $3H_{\rm kick} = m_{S, \rm kick} \equiv \sqrt{V'(S_{\rm ini})/S_{\rm ini}}$. This happens at the Hubble scale, $H_{\rm kick} =  \sqrt{2 \lambda_H}H_{\rm inf}/3,$
which is slightly after inflation ends.
To prevent the `second inflation' (i.e. the scalar field dominates the Universe before the kick occurs), we require that $\rho_{\chi, \rm ini} < \rho_{\rm kick}^{\rm rad} = 3 M_{\rm Pl}^2 H_{\rm kick}^2$ with $M_{\rm pl} \simeq 2.44\times 10^{18}~\rm GeV$, which translates into the lower bound of $\lambda$,
\begin{equation}
    \lambda > \frac{3\lambda_H}{8} \left(\frac{H_{\rm inf}}{\MPl}\right)^2 \simeq 2.3 \cdot 10^{-10}  \left(\frac{\sqrt{\lambda_H} H_{\rm inf}}{6 \cdot 10^{13} ~ {\rm GeV}}\right)^2.
    \label{eq:second_inflation_bound}
\end{equation}

As the field orbits inside a quartic potential, the total energy density of the complex field redshifts like radiation as $\sim a^{-4}$.
The last remaining ingredient \emph{iv)} `the radial damping' must occur before the field reaches the potential minimum. 
We refer to e.g., \cite{Co:2019wyp, Co:2020jtv, Co:2021lkc, Gouttenoire:2021jhk, Eroncel:2024rpe} for the realizations of radial damping using the interactions between the complex field and other particles in the thermal plasma.
Once the orbit reaches the minimum $S = f_a$, the axion can be considered as a freely rolling field, whose fluctuations can be efficiently induced by the curvature perturbation. 

From the time of the kick to the time when the complex field reaches the bottom of its potential,
the Universe has expanded by ${a_{\rm kin}}/{a_{\rm kick}} = {S_{\rm ini}}/{f_a} = \sqrt{1 + \beta^2}.$
The energy densities in the rotating homogeneous axion $\overline{\phi}$ and the total energy density of the Universe at the start of the kination behavior read,
\begin{align}
    \overline{\rho}_{\phi, \rm kin} &= \epsilon \rho_{\chi, \rm ini} \left(\frac{a_{\rm kick}}{a_{\rm kin}}\right)^4 = \epsilon \left(\frac{\lambda_H^2}{4\lambda}\right) \frac{H_{\rm inf}^4}{(1+\beta^2)^2},
    \label{eq:model_rotation_energy_KD}\\
    \rho_{\rm tot, kin} &= \rho_{\rm kick}^{\rm rad} \left(\frac{a_{\rm kick}}{a_{\rm kin}}\right)^4 = \left(\frac{2 \lambda_H}{3}\right)\frac{\MPl^2 H_{\rm inf}^2}{(1+\beta^2)^2},
    \label{eq:model_background_KD}
\end{align}
where we use that both quantities scale as $a^{-4}$, and $\epsilon$ parametrizes the energy fraction in the axion compared to that of the radial mode.
$\epsilon$ encodes the orbit ellipticity and can be determined by the exact form of the explicit $U(1)$--breaking term, see e.g.~\cite{Co:2020dya,Gouttenoire:2021jhk}. 
Since $\Omega_{\rm fluct}^0/\Omega_{\rm zero}^0 \propto \sqrt{\epsilon}\,\mathcal{P}_\mathcal{R}(k_{\rm kin})$, a smaller $\epsilon$ can be compensated by a larger $\mathcal{P}_\mathcal{R}$.
We assume $\epsilon \sim 1$, as for $\epsilon \ll 1$, the complex field could also experience parametric resonance \cite{Co:2020dya}. Note that $\epsilon \sim 1$ can be obtained in supersymmetric realizations where the potential is nearly-quadratic \cite{Co:2019wyp,Gouttenoire:2021jhk,Eroncel:2024rpe}.

Using Eqs.~\eqref{eq:model_rotation_energy_KD}--\eqref{eq:model_background_KD} in Eq.~\eqref{eq:yield_ratio},  we show in Fig.~\ref{fig:uv_complete} the region where DM comes from the curvature-induced fluctuation and the kinetic misalignment, $\Omega_{\rm DM}^0 = \Omega_{\rm fluct}^0 + \Omega_{\rm zero}^0$ where $\Omega_{\rm DM}^0 = 0.2657$ \cite{ParticleDataGroup:2024cfk} is the fractional energy density in DM today and we assume no interaction between the fluctuation and zero modes.
For  ${\mathcal{P}_{\mathcal{R}}(k)}$ from Eq.~\eqref{eq:Delta_R_k}, the curvature-induced axion contributes to DM with at least more than $1\%$ of the zero-mode abundance, which goes up to more than $ 10\%$ for the QCD axion.
The purple region is ruled out by the second inflation constraint \eqref{eq:second_inflation_bound}.
The curvature-induced fluctuation opens up the model parameter space.
Note also the dependence on the inflationary scale $H_{\rm inf}$ and the Hubble-mass coefficient $\lambda_H$ in  Eqs.~\eqref{eq:model_rotation_energy_KD}-\eqref{eq:model_background_KD} such that [Eq.~\eqref{eq:yield_ratio}] scales as $\Omega_{\rm fluct}^0/\Omega_{\rm zero}^0 \propto \sqrt{\lambda_H}H_{\rm inf}$. Fig.~\ref{fig:uv_complete} uses the largest value suggested by Planck data \cite{Planck:2018jri} and a fixed $\lambda_H = 1$.
In the bottom panel, we show predictions for  $\mathcal{P}_{\mathcal{R}}(k_{\rm kin})=10^{-3}$, close to the primordial black holes overproduction bound \cite{Gow:2020bzo}, that leads to more enhanced axion fluctuations and larger parameter space.
We have checked that, for $m_a <100\,{\rm eV}$, the curvature-induced fluctuation dominates the DM abundance if $\mathcal{P}_{\mathcal{R}}(k_{\rm kin}) \gtrsim 3.5\times 10^{-7}$. We show the parameter spaces for other $\mathcal{P}_{\mathcal{R}}(k_{\rm kin})$ values in Supplemental Material \cite{SupplementalMaterial}.
For the parameter space in Fig.~\ref{fig:uv_complete}, the temperature at which the fluctuation (of $k_{\rm kin}$ mode) becomes non-relativistic is $T_{\rm NR} \gtrsim \textrm{MeV}$.

\textit{Experimental tests.}--In our example model, the DM from kinetic misalignment alone cannot explain DM for $m_a \lesssim 2.1\,{\rm meV}$ due to the second inflation bound \eqref{eq:second_inflation_bound}. As shown in Fig.~\ref{fig:uv_complete}, the new mechanism opens up this range of $m_a$, which is a target region for the ALPS II experiment \cite{Ortiz:2020tgs}, for helioscopes BabyIAXO and  IAXO \cite{Shilon_2013,Armengaud:2014gea}, and for future haloscopes \cite{AxionLimits}. 
Regarding gravitational signatures, it was shown in \cite{Eroncel:2022efc} that kinetic fragmentation of the axion condensate can lead to compact mini-halos. A new analysis will be required to study the formation of compact structures in the new scenario studied in this letter. 
Next, we also stress the potential signatures in gravitational waves, which will arise in the regime where the field temporarily dominates the energy density of the Universe over radiation, see Fig.~\ref{fig:kination_domination}.

\textit{From radiation-domination to kination}--If the fast-rolling axion dominates the Universe, it can lead to a kination era if the quartic potential (\ref{eq:complex_scalar_potential}) is replaced by a nearly-quadratic one \cite{Co:2019wyp,Co:2020jtv,Co:2021lkc,Gouttenoire:2021wzu,Gouttenoire:2021jhk}:
\begin{equation}
\label{quadraticpotential}
    V(\chi) = m_r^2 |\chi|^2\left(\log\frac{|\chi|^2}{f_a^2} - 1 \right) + m_r^2 f_a^2 + \frac{\kappa^2}{M_{\rm Pl}^{2l-6}}|\chi|^{2l-2},
 \end{equation}
together with the Hubble-mass term in Eq.~\eqref{eq:complex_scalar_potential}.
Its cosmological history is reviewed exhaustively in \cite{Gouttenoire:2021jhk}.
With the same ingredients as for the quartic model, 
the fluctuations and the zero mode can produce axion DM over a large region of parameter space shown in Fig.~\ref{fig:kination_domination} (see Supplemental Material \cite{SupplementalMaterial} for the slightly different relic abundance expressions).
Strikingly, the kination era induced by this DM axion imprints a triangular smoking-gun signature \cite{Co:2021lkc,Gouttenoire:2021wzu,Gouttenoire:2021jhk} in the inflationary gravitational-wave background that could be observed at LISA \cite{LISACosmologyWorkingGroup:2022jok}, ET \cite{ET:2019dnz} and BBO \cite{Yagi:2011wg}.
Note that our results rely on the fluctuations produced during the kination era only. The effect of the matter era preceding kination could, in fact, enhance the relic abundance, which we leave for a future analysis.

\textit{Conclusion.}--We presented a novel source for scalar DM in models where the scalar field experienced a short stage of fast-rolling in its early history.
We studied the implications in the framework of rotating axion models. Assumptions and constraints on parameters are similar to models of kinetic misalignment, in particular those conditions related to radial mode damping as studied extensively in \cite{Gouttenoire:2021jhk} and \cite{Eroncel:2024rpe}.
What we learn in this letter is that  axion DM can be dominantly produced from primordial curvature perturbations, as fluctuations rather than as the well-known oscillations of the zero-mode condensate, 
without demanding any specific assumption on the realisation of inflation.

Our findings are highly relevant to a large number of upcoming axion experiments that will probe the specific region of parameter space (where the axion cannot be DM from the standard misalignment mechanism \cite{Preskill:1982cy,Abbott:1982af,Dine:1982ah}) where we predict our novel source of axions to be the dominant one and sufficient to explain DM. Precise predictions
will need to be checked with numerical simulations because of backreaction effects. While fluctuations from the condensate fragmentation in \cite{Eroncel:2022vjg} do not impact much the axion relic abundance because they are produced when mostly non-relativistic, in this work, the fluctuations are highly relativistic at the time of production and interactions between the fluctuations and the background could modify the predictions presented in our analysis. The corresponding lattice simulations will be challenging as they involve a long evolution and a large hierarchy of scales.

\vspace{0.5em}

\textit{Acknowledgements.}--We thank Arushi Bodas, Akshay Ghalsasi, Keisuke Harigaya, Raymond Co, and Lian-Tao Wang for informing us of their upcoming related paper \cite{Bodas:2025eca} while this work was under completion.
YG acknowledges support by the Cluster of Excellence ``PRISMA+'' funded by the German Research Foundation (DFG) within the German Excellence Strategy (Project No. 390831469), and by a fellowship awarded by the Azrieli Foundation.
PS is supported by Generalitat Valenciana Grants:  PROMETEO/2021/083 and CIPROM/2022/69. This work is also supported by the Deutsche Forschungsgemeinschaft under Germany’s Excellence Strategy---EXC 2121 ``Quantum Universe"---390833306.
The work of RS is supported in part by JSPS KAKENHI Grant Numbers~23K03415, 24H02236, and 24H02244. This article/publication is based upon work from COST Action COSMIC WISPers CA21106, supported by COST (European Cooperation in Science and Technology).

\appendix
\onecolumngrid

\fontsize{11}{13}\selectfont

\newpage
\vspace{1cm}

\begin{center}
\textbf{\Large Supplemental Material
}\\[0.5em]
{\large Cem Er\"{o}ncel, Yann Gouttenoire, Ryosuke Sato, G\'{e}raldine Servant, Peera Simakachorn}
\end{center}
\noindent

\renewcommand{\tocname}{\large  Contents
\vspace{-0.2 cm}}

   \titleformat{\section}
  {\normalfont\fontsize{12}{14}\bfseries  \centering }{\thesection:}{1em}{}
  \titleformat{\subsection}
  {\normalfont\fontsize{12}{14}\bfseries \centering}{\thesubsection.}{1em}{}
  \titleformat{\subsubsection}
  {\normalfont\fontsize{12}{14}\bfseries \centering}{\thesubsubsection)}{1em}{}
  
     \titleformat{\paragraph}
  {\normalfont\fontsize{12}{14}\bfseries  }{\thesection:}{1em}{}

   \counterwithin{figure}{section}
    \renewcommand\thefigure{S\arabic{figure}} 
 
{
  \hypersetup{linkcolor=black}

}

\section{Derivation of $\rho_{\rm fluct}$}
\label{app:derivation_rho_fluct}

In this Appendix, we derive the energy density of the scalar field fluctuations and describe its evolution from the beginning of the fast-rolling regime until today. The sub-sections~\ref{sec:eom-general} and~\ref{sec:fluctuations-fast-rolling} review the derivation already presented in~\cite{Eroncel:2022vjg} while the results derived in sub-sections~\ref{sec:fluctuations-non-perturbative} and~\ref{sec:fluctuations-energy} are new.

\subsection{Equations of Motion (EoM) for the background and fluctuations}
\label{sec:eom-general}

Our goal is to study the effect of curvature fluctuation $\mathcal{R}_\mathbf{k}$ on a scalar field which is decomposed as
\begin{equation}
    \label{eq:scalar-field-decomposition}
    \phi(\eta,\vb{x})=\overline{\phi}(\eta)+\delta\phi(\eta,\vb{x})=\overline{\phi}(\eta)+\int\frac{\dd[3]{\vb{k}}}{(2\pi)^3}\delta\phi_{\mathbf k}(\eta)e^{-i \vb{k}\cdot\vb{x}},
\end{equation}
where $\overline{\phi}$ and $\delta\phi$ denote the background and the fluctuations respectively. Denoting by $V(\phi)$ its potential, the equation of motion (EoM) for the background is
\begin{equation}
    \label{eq:EOM_scalar_bg}
    \overline{\phi}''+2\mathcal{H}\overline{\phi}'+a^2 \pdv{V}{\phi}=0,
\end{equation}
while the EoM for the fluctuation Fourier modes $\delta \phi_{\mathbf{k}}$ in the limit $\Phi \ll 1$ and $\delta\phi \ll 1$ read~\cite{Riotto:2002yw}
\begin{equation}
\label{eq:EOM_scalar}
\delta\phi_{\mathbf{k}}''+2\mathcal{H}\delta\phi_{\mathbf{k}}' +\qty(k^2+a^2\pdv[2]{V}{\phi})\delta\phi_{\mathbf{k}}  = - 2 \Phi_{\mathbf{k}} \pdv{V}{\phi} +  4 \Phi_{\mathbf{k}}'\overline{\phi}',
\end{equation}
In both equations, the potential derivatives are evaluated at $\phi=\overline{\phi}$. The solution of $\Phi_{\mathbf k}$ during radiation ($\omega = 1/3$) is:~\cite{Mukhanov:2005sc}
\begin{equation}
\label{eq:Phik_formula}
    \Phi_{\mathbf{k}}(k\eta) =\!\frac{9}{\sqrt{3}}\frac{\Phi_{\mathbf{k}}(0)}{k\eta}j_1\!\!\left(\frac{k\eta}{\sqrt{3}}\right)\! \xrightarrow[k\eta \gg 1]{}6\mathcal{R}_{\mathbf{k}}(0)\frac{\cos(k\eta/\sqrt{3})}{(k\eta)^{2}},
\end{equation}
where $j_1(x)$ is the spherical Bessel function and $\mathcal{R}(0) = 3\Phi(0)/2$ is the comoving curvature perturbation on super-horizon scales in radiation era~\cite{Mukhanov:2005sc}.

\subsection{Fluctuations in the fast-rolling regime}
\label{sec:fluctuations-fast-rolling}
We assume the scalar field undergoes a period of fast-rolling during which the potential $V(\phi)$ is negligible compared to its
kinetic and gradient energies. In this regime, the EoM \eqref{eq:EOM_scalar} simplifies to
\begin{equation}
\label{eq:EOM_scalar_early}
\delta\phi_{\mathbf{k}}''+2\mathcal{H}\delta\phi_{\mathbf{k}}' +k^2\delta\phi_{\mathbf{k}}=  4 \Phi_{\mathbf{k}}'\overline{\phi}'.
\end{equation}
The fluctuation $\delta\phi_{\mathbf k}$ is sourced by $\Phi_{\mathbf k}$ around the time of the horizon re-entry ($k = \mathcal{H}=aH$).

The solution of Eq.~\eqref{eq:EOM_scalar_early} is derived in Appendix B of~\cite{Eroncel:2022vjg} for the modes that enter the horizon during the fast-rolling period by assuming adiabatic initial conditions at $a_{\rm kin}$. In the sub-horizon limit $k/\mathcal{H}\gg 1$, the solution is approximated by
\begin{equation}
    \label{eq:mode-function-subhorizon}
        \delta\phi_{\mathbf k}\approx \alpha \mathcal{R}_{\mathbf k}(0)\eval{\frac{\overline{\phi}'}{\mathcal{H}}}_{\rm kin}\qty(\frac{a_{\rm kin}}{a})\cos\qty(\frac{k}{\mathcal{H}}),
\end{equation}
where
\begin{equation}
    \label{eq:alpha-def}
    \alpha \equiv \frac{2}{3}\qty[\sqrt{3}\ln\qty(\frac{\sqrt{3}+1}{\sqrt{3}-1})-\frac{3}{2}]\simeq 0.52.
\end{equation}
The scalar field density contrast in the sub-horizon limit is
\begin{equation}
    \label{eq:density-contrast-early-subhorizon}
    \delta_{\phi,\mathbf k}=2\qty(\frac{\delta\phi_{\mathbf k}'}{\overline{\phi}'}-\Phi_{\mathbf k})\simeq 2\frac{\delta\phi_{\mathbf k}'}{\overline{\phi}'}\simeq -2\alpha \mathcal{R}_{\mathbf k}(0)\qty(\frac{k}{\mathcal{H}})\sin\qty(\frac{k}{\mathcal{H}})\propto a.
\end{equation}
We see that the density contrast grows linearly with the scale factor. From this expression, we can obtain the dimensionless power spectrum of the scalar field density fluctuations as
\begin{equation}
    \label{eq:power-spectrum-early}
    \mathcal{P}_{\phi}(k;\eta)\simeq 4\alpha^2 \mathcal{P}_{\mathcal{R}}(k)\qty(\frac{k}{\mathcal{H}})^2\sin^2\qty(\frac{k}{\mathcal{H}})\to 2\alpha^2 \mathcal{P}_{\mathcal{R}}(k)\qty(\frac{k}{\mathcal{H}})^2,
\end{equation}
where in the last step we took a time average of the mode function oscillations. This expression implies that the mode with the wavenumber $k$ becomes non-perturbative when
\begin{equation}
    \label{eq:non-perturbative-mode}
    \mathcal{P}_{\phi}(k;\eta)>1 \implies \qty(\frac{a}{a_{\rm kin}})^2 >\frac{1}{2\alpha^2}\mathcal{P}_{\mathcal{R}}^{-1}(k)\qty(\frac{k}{\mathcal{H}_{\rm kin}})^{-1}.
\end{equation}
We also see that the mode with the highest wavenumber becomes non-perturbative first. Since the analysis presented in this section is valid for the modes that enter the horizon during the fast-rolling regime, the first mode that becomes non-perturbative is the mode for which $k=\mathcal{H}_{\rm kin}$. Finally, we conclude that the linear perturbation theory breaks down when
\begin{equation}
    \label{eq:non-perturbative-system}
    \mathcal{P}_{\phi}(k;\eta_{\rm kin})>1 \implies \qty(\frac{a}{a_{\rm kin}})^2 >\frac{1}{2\alpha^2}\mathcal{P}_{\mathcal{R}}^{-1}(k_{\rm kin}).
\end{equation}

\subsection{Fluctuations in the non-perturbative regime}
\label{sec:fluctuations-non-perturbative}

We now discuss the fate of the scalar field fluctuations in the "non-perturbative regime", i.e. when the inequality given in Eq.~\eqref{eq:non-perturbative-system} is satisfied. In this section we will also allow that the scalar field potential has a, in general time-dependent, mass term so that they can be dark matter. We will be interested in the modes that are deep inside the horizon, but were super-horizon at the beginning of the fast-rolling regime. For these modes, the effect of the curvature perturbations is negligible, so that the EoM for the scalar field becomes
\begin{equation}
    \label{eq:scalar-field-general}
    \phi''+2\mathcal{H}-\nabla^2 \phi+V'(\phi)=0.
\end{equation}
We can again decompose the scalar field as in Eq.~\eqref{eq:scalar-field-decomposition}, but now we shouldn't suppose that $\delta\phi\ll \overline{\phi}$. If the scalar field potential $V(\phi)$ is purely a mass term, i.e. $V(\phi)=\frac{1}{2}m^2(\eta)\phi^2$, then the EoMs for the background and the fluctuations decouple, and we can express them as
\begin{equation}
    \label{eq:bg-eom-mass}
    \overline{\phi}''+2\mathcal{H}\overline{\phi}'+m^2(\eta)\overline{\phi}=0,
\end{equation}
and
\begin{equation}
    \label{eq:fluct-eom-mass}
    \delta\phi_{\mathbf k}''+2\mathcal{H}\delta\phi_{\mathbf k}'+\qty[k^2+a^2m^2(\eta)]\delta\phi_{\mathbf k}=0,
\end{equation}
respectively. We note that such a separation cannot be done for a generic potential and we will need lattice simulations to study the system. This is left for a separate study.

If we assume that the Hubble scale $H$ and the scalar field mass $m(\eta)$ vary slowly compared to $\Upsilon_k(t)=\sqrt{k^2/a^2(t)+m^2(t)}$, then we can use the WKB approximation to express the solution of the mode function in physical time as
\begin{equation}
    \label{eq:eom-fluct-late-physical-time}
    \delta\phi_{\mathbf k}=\frac{c_{\mathbf k}}{a^{3/2}(t)\sqrt{\Upsilon_k(t)}}\cos\qty(\int^{t} \dd{t}'\Upsilon_k(t')).
\end{equation}
The coefficient $c_{\mathbf k}$ can be determined by matching this solution to Eq.~\eqref{eq:mode-function-subhorizon} in the relativistic limit $k/a(t)\gg m(t)$. This procedure fixes $c_{\mathbf k}$ to be
\begin{equation}
    \label{eq:ck-factor}
    c_{\mathbf k}=\alpha \mathcal{R}_{\mathbf k}(0)\sqrt{k}\,a_{\rm kin}\eval{\frac{\dot{\overline{\phi}}}{H}}_{\rm kin}
\end{equation}
In the next section, we will use this result to calculate the energy density stored in the fluctuations. 

\subsection{Energy density stored in the fluctuations}
\label{sec:fluctuations-energy}
In this section, we will calculate the energy density stored in the fluctuations. Starting with the stress-energy tensor of a scalar field
\begin{equation}
    \label{eq:scalar-field-stress-energy}
    T_{\mu\nu}=\partial_{\mu}\phi\partial_{\nu}\phi-g_{\mu\nu}\qty[\frac{1}{2}g^{\alpha\beta}\partial_{\alpha}\phi\partial_{\beta}\phi+V(\phi)]
\end{equation}
and using the metric given in the manuscript, we can calculate the energy density of the scalar field as
\begin{equation}
    \label{eq:scalar-field-rho-general}
    \rho_{\phi}=-T^0_{\;0}=\frac{1}{2}\qty(1-2\Phi)\frac{\phi'^2}{a^2}+\frac{1}{2}\qty(1+2\Phi)\qty(\frac{\grad{\phi}\cdot\grad{\phi}}{a^2})+V(\phi).
\end{equation}
In deriving this expression we have neglected the terms on the order of $\Phi^2$, but the scalar field is kept non-perturbative. By substituting the expansion in Eq.~\eqref{eq:scalar-field-decomposition} into Eq.~\eqref{eq:scalar-field-rho-general} and assuming a quadratic potential, we can show that
\begin{equation}
    \label{eq:scalar-field-rho-decomposition}
    \rho_{\phi}=\rho_{\phi}^{(0)}+\rho_{\phi}^{(1)}+\rho_{\phi}^{(2)}+\rho_{\phi}^{(3)},
\end{equation}
where
\begin{align}
    \rho_{\phi}^{(0)}&=\frac{1}{2}\frac{\overline{\phi}'^2}{a^2}+V(\overline{\phi})=\overline{\rho_{\phi}},\\
    \rho_{\phi}^{(1)}&=\frac{\overline{\phi}'\delta\phi'}{a^2}-\Phi\overline{\phi}'^2+m^2\overline{\phi}\delta\phi\\
    \rho_{\phi}^{(2)}&=\frac{1}{2}\frac{\delta\phi'^2}{a^2}+\frac{1}{2}\frac{\grad{\delta\phi}\cdot\grad{\delta\phi}}{a^2}+\frac{1}{2}m^2\delta\phi^2\\
    \rho_{\phi}^{(3)}&=-\Phi\frac{\delta\phi'^2}{a^2}+\Phi\frac{\grad{\delta\phi}\cdot\grad{\delta\phi}}{a^2},
\end{align}
with the superscript denoting the order of the perturbation. For the dark matter density coming from the fluctuations, what matters is the average of the energy density over a sufficiently large cosmological volume. Since the scalar field fluctuations $\delta\phi$ is sourced by the curvature perturbations $\Phi$ which are Gaussian, the volume average of the first order $\rho_{\phi}^{(1)}$ as well as the third order $\rho_{\phi}^{(3)}$ contributions vanish. Thus, we can define the energy density stored in the fluctuations as
\begin{equation}
    \label{eq:rho-fluct}
    \rho_{\rm fluct}\equiv \expval{\rho_{\phi}}-\overline{\rho_{\phi}}=\expval{\rho^{(2)}_{\phi}}=\frac{1}{2}\int \frac{\dd[3]{\mathbf k}}{(2\pi)^3}\qty(\abs{\delta\dot{\phi}_{\mathbf k}(t)}^2+\Upsilon_k^2(t)\abs{\delta\phi_{\mathbf k}(t)}^2).
\end{equation}
For these modes, we can use Eq.~\eqref{eq:mode-function-subhorizon} as the solution of the mode functions. By plugging Eq.~\eqref{eq:eom-fluct-late-physical-time} into \eqref{eq:rho-fluct} and using Eq.~\eqref{eq:ck-factor} we obtain
\begin{equation}
    \label{eq:rho-fluct-integral}
    \rho_{\rm fluct}(t)=\alpha^2\frac{\dot{\phi}_{\rm kin}^2}{2}\qty(\frac{a_{\rm kin}}{a(t)})^3\int_{\mathcal{H}_{\rm nl}}^{\mathcal{H}_{\rm kin}}\dd{\log k}\qty(\frac{k}{\mathcal{H}_{\rm kin}})\mathcal{P}_{\mathcal{R}}(k)\sqrt{\qty(\frac{k}{\mathcal{H}_{\rm kin}})^2 \qty(\frac{a_{\rm kin}}{a(t)})^2 + \qty(\frac{m(t)}{H_{\rm kin}})^2}
\end{equation}
This expression is consistent with Eq.~\eqref{eq:rho_fluct_end_1} in the fast-rolling regime where all the modes are relativistic, and with Eq.~\eqref{eq:rho_fluct_end_2} at late times when all the modes are non-relativistic. We can observe that the dominant contribution to the energy density comes from the mode with the highest wavenumber, i.e. $k_{\rm kin}$. By approximating $\mathcal{P}_{\mathcal{R}}(k)\simeq \mathcal{P}_{\mathcal{R}}(k_{\rm kin})$ we can evaluate the integral to show that the energy density in the fluctuations behave as
\begin{equation}
    \label{eq:rho-fluct-early-late}
    \rho_{\rm fluct}(a)\simeq\begin{cases}
        \dfrac{\alpha^2}{4}\mathcal{P}_{\mathcal{R}}(k_{\rm kin})\dot{\overline{\phi}}_{\rm kin}^2\qty(\dfrac{a_{\rm kin}}{a})^4,& \dfrac{k_{\rm kin}}{a}\gg m(a)\\[0.65em]
        \dfrac{\alpha^2}{2}\mathcal{P}_{\mathcal{R}}(k_{\rm kin})\dot{\overline{\phi}}_{\rm kin}^2\qty(\dfrac{m_0}{H_{\rm kin}})\qty(\dfrac{a_{\rm kin}}{a})^3,& \dfrac{k_{\rm kin}}{a}\ll m(a)
    \end{cases}
\end{equation}
We see that the energy density in fluctuations redshifts as radiation when the dominant mode $k_{\rm kin}$ is relativistic, and redshifts as matter when the dominant mode is non-relativistic.

\section{Equation of state of the fluctuations}
\label{app:equation_of_state}
The equation of the state (EoS) of the fluctuations can be defined by $\omega_{\rm fluct} = \langle p_{\rm fluct} \rangle/\langle \rho_{\rm fluct} \rangle$ where $\rho_{\rm fluct}=\rho_\phi - \overline{\rho}_\phi$ is the energy density in Eq.~\eqref{eq:scalar-field-rho-general} and $p_{\rm fluct} = p_\phi - \overline{p}_\phi$ is the pressure defined as 
\begin{equation}
    \label{eq:scalar-field-pressure-general}
    p_{\phi}=\frac{1}{2}\qty(1-2\Phi)\frac{\phi'^2}{a^2}-\frac{1}{6}\qty(1+2\Phi)\qty(\frac{\grad{\phi}\cdot\grad{\phi}}{a^2})-V(\phi).
\end{equation}
In terms of the Fourier mode $\delta \phi_k$ and assuming $V(\phi) = m_a^2\phi^2/2$, the EoS reads
\begin{equation}
    \omega_{\rm fluct} = \frac{\left\langle\int d^3k \left[\dot{\delta\phi^2_k} - \left(\frac{k^2}{3a^2} + m_a^2 \right)\delta\phi^2_k\right]\right\rangle}{\left\langle\int d^3k \left[\dot{\delta\phi^2_k} + \left(\frac{k^2}{a^2} + m_a^2 \right)\delta\phi^2_k\right]\right\rangle},
\end{equation}
where the integration runs from $k_{\rm osc}$ to $k_{\rm kin}$.
Using the $\delta\phi_k$-solution \eqref{eq:eom-fluct-late-physical-time}, the EoS can be written as 
\begin{align}
    \omega_{\rm fluct}(a) \simeq \frac{\int d k \mathcal{P}_\mathcal{R}(k) \Upsilon_k \left\{1-{\Upsilon_k^{-2}}\cdot\left[{k^2/(3a^2) +m_a^2}\right]\right\}}{2\int d k \mathcal{P}_\mathcal{R}(k) \Upsilon_k}.
    \label{eq:EoS_general}
\end{align}
In the relativistic limit $k/a \gg m_a$, the fluctuations behave as radiation, $\omega_{\rm fluct} \to 1/3$, while $\omega_{\rm fluct} \to 0$ in the non-relativistic limit $k/a \ll m_a$.

Since we know that the dominant energy density of fluctuations is generated around the scale $k_{\rm kin}$, let us assume $\mathcal{P}_{\mathcal{R}}(k)= A_s \delta(k-k_{\rm kin})$ which simplifies our calculation to,
\begin{align}
    \omega_{\rm fluct}(a) \simeq \left[3(1+m_a^2a^2/k_{\rm kin}^2)\right]^{-1}.
    \label{eq:eos_simplified_KD}
\end{align}
The EoS matches the 1/3 behavior in the relativistic limit and decreases with $a^{-2}$ in the non-relativistic limit.
Using that $k_{\rm kin}/a_{\rm kin} = H_{\rm kin} = (\pi/\sqrt{90})g_*^{1/2}(T_{\rm kin}) T_{\rm kin}^2/M_{\rm Pl}$, the EoS of fluctuations at matter-radiation equality is,
\begin{align}
    \omega_{\rm fluct,eq} \simeq \frac{\pi^2 g_*(T_{\rm kin})}{270}\left[\frac{g_{*s}(T_{\rm eq})}{g_{*s}(T_{\rm kin})}\right]^{2/3}\frac{T_{\rm eq}^2 T_{\rm kin}^2}{m_a^2 M_{\rm Pl}^2},
    \label{eq:EoS_today_KD_mode}
\end{align}
where the derivation also uses that $\omega_{\rm fluct,\rm eq}\ll 1$.
The bound on the warmness of DM discussed in the main text is thus translated into the upper bound on $T_{\rm kin}$ [Eq.~\eqref{eq:warmness_bound}].
If we do the same analysis as in Eq.~\eqref{eq:eos_simplified_KD} for other fluctuation modes, it is easy to show that $\omega_{\rm fluct,\rm eq}(k<k_{\rm kin}) < \omega_{\rm fluct,\rm eq}(k_{\rm kin})$.

In the case where the scalar field dominates the Universe, the bound from warmness of DM in Eq.~\eqref{eq:warmness_bound}, which assumes the scalar field is sub-dominant, is changed. Using Eq.~\eqref{eq:eos_simplified_KD} with $k_{\rm kin}/a_{\rm kin} = H_{\rm kin} = \overline{\rho}_{\phi,\rm kin}^{1/2}/(\sqrt{3}M_{\rm Pl})$ where $\overline{\rho}_{\phi,\rm kin}>\rho_{\rm rad,kin}$, the warmness bound in the case of kination domination reads,
\begin{align}
    \omega_{\rm fluct} \simeq \left[\frac{g_{*s}(T_{\rm eq})}{g_{*s}(T_{\rm kin})}\right]^{2/3}\frac{T_{\rm eq}^2 \,\overline{\rho}_{\phi,\rm kin}}{m_a^2 M_{\rm Pl}^2 T_{\rm kin}^2}.
\end{align}

\section{An estimate of the backreaction}
\label{App:backreaction}
Throughout this work we have neglected interactions between the fluctuations and the homogeneous mode by approximating the scalar field potential by a quadratic one. However, in physically motivated scenarios a small mass for the scalar field is generated by the non-perturbative effects which implies a periodic potential for the scalar field. In particular, if the rolling scalar field is an axion, then its potential is given by
\begin{equation}
    \label{eq:axion-potential}
    V(\phi,T)=m^2_a(T)f_{a}^2\qty[1-\cos\qty(\frac{\phi}{f_{a}})].
\end{equation}
We now discuss whether this potential affects the results presented in this letter.

During the fast-rolling regime, the scalar field is dominated by its kinetic energy so the results we derived in the second section of the main text and in Appendix~\ref{sec:fluctuations-fast-rolling} are not affected. The self-interactions due to the non-quadratic potential might be prominent once the size of the potential becomes comparable to the kinetic energy of the homogeneous mode or to the gradient energy stored in the fluctuations. These non-linear interactions can have two effects: First, they can cause the fluctuations to backreact on the homogeneous mode, altering its evolution; second, they can yield a modification of the fluctuation power spectrum due to the interactions between the modes. In this section we focus on the former effect, leaving the latter case to a future study. 

The equation of motion for the scalar field with the potential given in Eq.~\eqref{eq:axion-potential} without any approximation reads
\begin{equation}
    \label{eq:scalar-general-eom}
    \ddot{\phi}+3H\dot{\phi}-\nabla^2\phi+m^2_a(t)f_{a} \sin\qty(\frac{\phi}{f_{a}})=0,
\end{equation}
We again expand the scalar field as the sum of the homogeneous mode and the fluctuations as in Eq.~\eqref{eq:scalar-field-decomposition}, plug into Eq.~\eqref{eq:scalar-general-eom} and take an ensemble average of the equation of motion. By using the fact $\expval{\delta\phi}=0$, this yields to
\begin{equation}
    \label{eq:scalar-general-eom-avg}
    \ddot{\overline{\phi}}+3H\dot{\overline{\phi}}+m^2_a(t)f_a\qty[\sin\qty(\frac{\overline{\phi}}{f_a}) + \sum_{n=2}^{\infty}\frac{1}{n!}\qty(\eval{\pdv[n]{}{(\phi/f_a)}\sin\frac{\phi}{f_{a}}}_{\phi=\overline{\phi}})\frac{\expval{(\delta\phi)^n}}{f_a^n}]= 0,
\end{equation}
We can perform the sum by assuming that the fluctuation modes $\delta\phi_{\mathbf k}$ obey the Gaussian statistics we find 
\begin{equation}
    \ddot{\overline{\phi}}+3H\dot{\overline{\phi}}+m^2_a(t)f_a\sin\qty(\frac{\overline{\phi}}{f_a})e^{-\delta_2}=0,\quad \delta_2=\frac{1}{2 f_{a}^2}\int\frac{\dd[3]{\mathbf k}}{(2\pi)^3}\abs{\delta\phi_{\mathbf k}}^2.
\end{equation}
We see that the effect of a large variance is the exponential suppression of the mass term. By using the mode function solution given in Eq.~\eqref{eq:eom-fluct-late-physical-time} together with Eq.~\eqref{eq:ck-factor}, we can calculate the variance as
\begin{equation}
    \label{eq:variance-result}
    \delta_2(a)=\frac{\alpha^2}{4}\frac{\dot{\overline{\phi}}_{\rm kin}^2}{f_a^2 H_{\rm kin}^2}\qty(\frac{a_{\rm kin}}{a})^3\int_{\mathcal{H}(a)}^{\mathcal{H}_{\rm kin}}\dd{k}\frac{\mathcal{P}_{\mathcal{R}}(k)}{\sqrt{\frac{k^2}{a^2}+m^2_a(a)}}.
\end{equation}
By approximating $\mathcal{P}_{\mathcal{R}}(k)\approx \mathcal{P}_{\mathcal{R}}(k_{\rm kin})$ in the relevant range we can perform the integral to get
\begin{equation}
    \label{eq:variance-result-2}
    \delta_2(a)=\frac{\alpha^2}{4}\frac{\dot{\overline{\phi}}_{\rm kin}^2}{f_a^2 H_{\rm kin}^2}\qty(\frac{a_{\rm kin}}{a})^2\mathcal{P}_{\mathcal{R}}(k_{\rm kin})\sinh^{-1}\eval{\qty(\frac{k/a}{m_a(a)})}^{\mathcal{H}_{\rm kin}}_{\mathcal{H}}.
\end{equation}
We can observe that the variance decreases with redshift so it is largest at early times. However, during these times, the potential is negligible anyway, so a large variance at early times does not alter the evolution of the homogeneous mode. The relevant question is whether this term can be large when the homogeneous mode begins to oscillate. If this happens, the oscillations of the zero mode would get delayed further and the zero mode would continue to redshift as kination, thus suppressing its contribution to the dark matter abundance today. 

By assuming that $k_{\rm kin}$ is relativistic at $a_{\rm osc}$, i.e. $(k/a_{\rm osc})/m(a_{\rm osc})\gg 1$, the variance becomes
\begin{equation}
    \label{eq:variance-result-3}
    \delta_2(a_{\rm osc})=\alpha^2\frac{m_{a, \rm osc}^2}{H_{\rm osc}^2}\mathcal{P}_{\mathcal{R}}(k_{\rm kin})\qty[\sinh^{-1}\qty(\frac{a_{\rm kin}}{a_{\rm osc}}\frac{H_{\rm kin}}{m_{a, \rm osc}})-\sinh^{-1}\qty(\frac{H_{\rm osc}}{m_{a, \rm osc}})],
\end{equation}
where we have used the fact that $\dot{\overline{\phi}}^2\propto a^{-6}$ between $a_{\rm kin}$ and $a_{\rm osc}$ and the fact that without the backreaction, oscillations of the homogeneous mode starts when $\overline{\rho}_{\phi,\rm kin}=\overline{\rho}_{\phi,\rm pot} \implies \dot{\overline{\phi}}^2_{\rm osc}=2 m_{a, \rm osc}^2 f_a^2$. If $k_{\rm kin}$ is relativistic at $a_{\rm osc}$, then the argument of the first $\sinh^{-1}$ function is large so we can approximate it as $\sinh^{-1}[{a_{\rm kin}}{H_{\rm kin}}/({a \,m_{a, \rm osc}})]\approx \log[2{a_{\rm kin} H_{\rm kin}}/({a\,m_{a, \rm osc}})]$. On the other hand, in the kinetic misalignment mechanism, the onset of oscillations is delayed compared to the standard misalignment, so $m_{a, \rm osc}\gg H_{\rm osc}$. Then, the second $\sinh^{-1}$ term can be approximated as $\sinh^{-1}(H_{\rm osc}/m_{a, \rm osc})\approx H_{\rm osc}/m_{a, \rm osc}$ so this term is negligible compared to the $\log$ term. So, our final result for the variance at the onset of zero-mode oscillations is
\begin{equation}
    \label{eq:variance-at-aosc}
    \delta_2(a_{\rm osc})\approx \alpha^2 \qty(\frac{m_{a, \rm osc}}{H_{\rm osc}})^2 \mathcal{P}_{\mathcal{R}}(k_{\rm kin})\log\qty(2\frac{a_{\rm kin}}{a}\frac{H_{\rm kin}}{m_{a, \rm osc}}).
\end{equation}
We expect that the onset of the zero-mode oscillations will be delayed due to the backreaction of the fluctuations if this term is larger than unity. Neglecting the $\log$ factor this occurs when
\begin{equation}
    \label{eq:backreaction-condition}
    \frac{m_{a, \rm osc}}{H_{\rm osc}}\gtrsim \frac{1}{\alpha \mathcal{P}_{\mathcal{R}}^{1/2}(k_{\rm kin})}.
\end{equation}
For the rotating axion model presented in the manuscript $m_{a, \rm osc}/H_{\rm osc}$ can be calculated for a given $m_a$ and $f_a$ as shown in Ref.~\cite{Eroncel:2022vjg}, and a large $m_{a, \rm osc}/H_{\rm osc}$ values correspond to the low $f_a$ region. As shown in Figures~\ref{fig:uv_complete},~\ref{fig:kination_domination}, and~\ref{fig:UV-complete_app}, in the low $f_a$ region dark matter is already dominated by the fluctuations so we do not expect that the backreaction modifies our conclusions in a qualitative way.

\section{More results on the rotating axion in the quartic potential}
\begin{figure}[h!]
    \centering
    \includegraphics[width=0.333\linewidth]{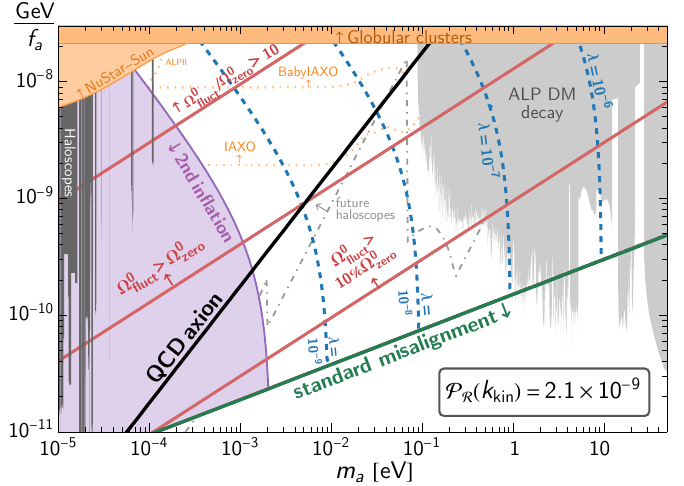}\hfill
    \includegraphics[width=0.333\linewidth]{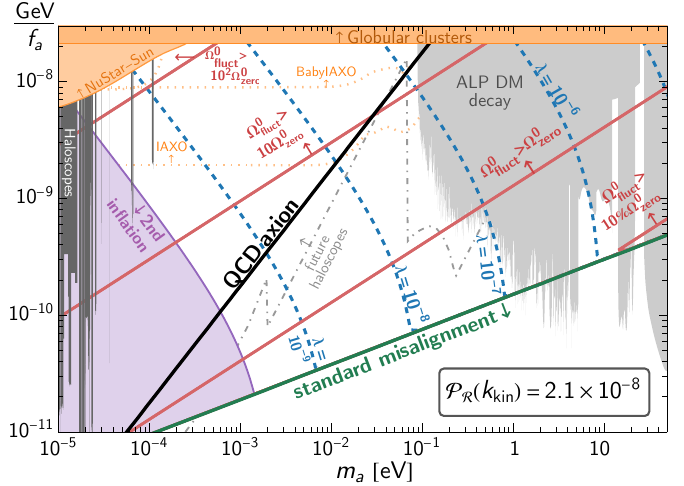}\hfill
    \includegraphics[width=0.333\linewidth]{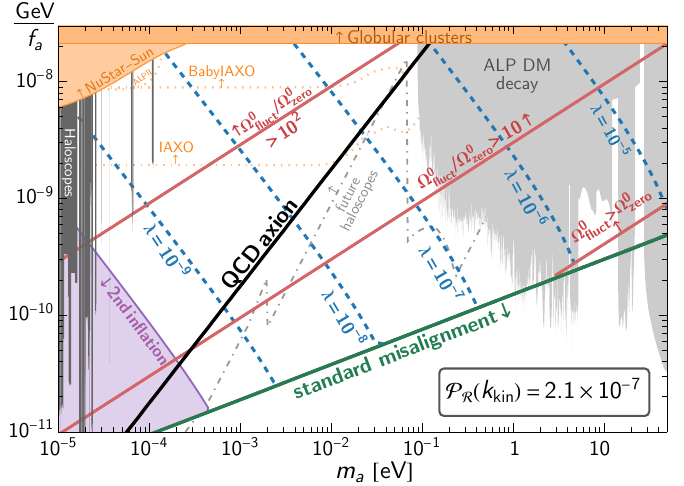}
    \caption{Same as Fig.~\ref{fig:uv_complete} of the main text with $H_{\rm inf} = 6 \times 10^{13}\,{\rm GeV}$, but with other values of $\mathcal{P}_{\mathcal{R}}(k_{\rm kin})$.}
    \label{fig:UV-complete_app}
\end{figure}

Fig.~\ref{fig:UV-complete_app} shows the axion parameter space, similar to Fig.~\ref{fig:uv_complete}, but for other values of $\mathcal{P}_{\mathcal{R}}(k_{\rm kin})$.
The larger $\mathcal{P}_{\mathcal{R}}(k_{\rm kin})$ increases the abundance of the curvature-induced DM and opens up more parameter space as the second inflation bound moves the lower $m_a$ region.
We also checked that for $\mathcal{P}_{\mathcal{R}}(k_{\rm kin}) \gtrsim 3.5 \times 10^{-7}$ the axion fluctuation dominates the DM energy density.

\section{
Dark matter abundance in the case where a kination era is triggered by the scalar field}
\label{app:yields_dominating_case}

In this appendix, we consider the scenario where the fast-rolling scalar dominates the energy density of the Universe and later becomes subdominant again. If it has an equation of state $\omega=1$, this will trigger a kination era. 
This is of high interest because it is associated with a large enhancement of the inflationary background of gravitational waves.

We estimate the scalar field abundance from the curvature-induced fluctuations and compare it to the background field abundance. We focus on the fluctuations generated during the fast-roll stage. However, the fluctuations generated during the preceding matter era would enhance the abundance; we will investigate this further in future work. 

\subsection{Model-independent expressions}

The kination era starts when the total energy density of the Universe is dominated by the energy density of scalar field  $\overline{\rho}_{\phi, \rm kin}$  (at $a_{\rm kin}$) and ends when the energy density drops to $\overline{\rho}_{\phi, \rm end}$ (at $a_{\rm end}$) below the energy density of the Standard Model radiation. The duration of the kination era is defined by the e-folding number of the cosmic expansion during the kination era, 
\begin{equation}
    N_{\rm KD} \equiv \log\left(\frac{a_{\rm end}}{a_{\rm kin}}\right) = \frac{1}{6}\log\left(\frac{\overline{\rho}_{\phi, \rm kin}}{\overline{\rho}_{\phi, \rm end}}\right).
\end{equation}
However, after the energy density of the scalar field becomes sub-dominant, it continues red-shifting as $a^{-6}$ in the radiation-dominated Universe until $a_{\rm osc}$.
The fast-rolling period of the scalar field goes through both kination and radiation eras, where the production of scalar fluctuations from the curvature perturbation is slightly different.

Let us split the fluctuations into two contributions, $\Omega_{\rm fluct}^{\rm tot} = \Omega_{\rm fluct}^{\rm I} + \Omega_{\rm fluct}^{\rm II}$, where $\Omega_{\rm fluct}^{\rm I}$  and $\Omega_{\rm fluct}^{\rm II}$ are the energy densities of fluctuations produced in the kination and radiation eras.
For $\Omega_{\rm fluct}^{\rm I}$, we adopt the fluctuation energy density which is generated during the kination era and is derived in Eqs.~(27) and (28) of \cite{Eroncel:2025bcb},
\begin{equation}
    \rho_{\rm fluct}^{\rm I}(\eta) = \frac{4}{\pi}\overline{\rho}_{\rm \phi} \int d\log k \, \left(\frac{k}{\mathcal{H}}\right) \mathcal{P}_{\mathcal{R}}(k).
\end{equation}
Using that $\mathcal{H} \propto a^{-2}$ during the kination era and performing a similar calculation as in the main text of this letter, the energy density of the non-relativistic radiation today reads, 
\begin{equation}
 \label{eq:Scalar_domination_rho_fluct_1}
    \Omega_{\rm fluct, 0}^{\rm I} = \frac{4}{\pi} \Omega_{\rm rad}^{0} \left(\frac{a_0 m}{\mathcal{H}_{\rm kin}}\right) \exp(2 N_{\rm KD}) \!\int_{k_{\rm end}}^{k_{\rm kin}} d\log k \, \mathcal{P}_{\mathcal{R}}(k).
\end{equation}
where we only integrate the fluctuations from curvature perturbations re-entering the horizon during the kination era ($k_{\rm kin}>k>k_{\rm end}$), and $\mathcal{H}_{\rm kin}/\mathcal{H}_{\rm kin} = \exp(2N_{\rm KD})$.
For the fluctuations produced during the radiation era after the kination era ends $a>a_{\rm end}$, we can use Eq.~\eqref{eq:rho_fluct_end_2} in the main text
 and replace `kin' with `end',
\begin{equation}
 \label{eq:Scalar_domination_rho_fluct_2}
    \Omega_{\rm fluct, 0}^{\rm II} = \alpha^2 \Omega_{\rm rad}^{0} \left(\frac{a_0 m}{\mathcal{H}_{\rm kin}}\right) \exp(2 N_{\rm KD}) \!\int_{k_{\rm osc}}^{k_{\rm end}}\frac{dk}{k}\left(\frac{k}{\mathcal{H}_{\rm end}}\right)\mathcal{P}_{\mathcal{R}}(k).
\end{equation}
where $\Omega_{\phi}(\eta_{\rm end}) = 1$.
For simplicity, assuming that  $\mathcal{P}_{\mathcal{R}}(k) \approx \mathcal{P}_{\mathcal{R}}(k_{\rm kin})$, we obtain
\begin{equation}
 \label{eq:Scalar_domination_rho_fluct_simple}
    \Omega_{\rm fluct, 0}^{\rm tot}= \Omega_{\rm fluct, 0}^{\rm I} + \Omega_{\rm fluct, 0}^{\rm II}  =  \Omega_{\rm rad}^{0} \left(\frac{a_0 m}{\mathcal{H}_{\rm kin}}\right) \exp(2 N_{\rm KD}) \mathcal{P}_{\mathcal{R}}(k_{\rm kin}) \left( \frac{8}{\pi} N_{\rm KD} + \alpha^2\right).
\end{equation}
where $k = \mathcal{H}$, $\log(k_{\rm kin}/k_{\rm end}) = 2N_{\rm KD}$, and $k_{\rm end} \gg k_{\rm osc}$ are used.
To connect to the model parameters which determine the cosmological history, we also provide the expression
\begin{align}
    \left(\frac{a_0 m}{\mathcal{H}_{\rm kin}}\right)\exp(2N_{\rm KD}) \simeq 1.49 \times 10^{13} \left(\frac{m}{1\,{\rm eV}}\right) \left(\frac{10^{9}\,{\rm GeV}}{\rho_{\rm kin}^{1/4}}\right) \left(\frac{g_{*s}^{1/3}(T_{\rm kin})}{g_{*s}^{1/4}(T_{\rm kin})}\right) \exp\left(\frac{3}{2}N_{\rm KD}\right).
\end{align}

We see that the abundance of fluctuations produced during the kination era is enhanced by a factor $N_{\rm KD}$  compared to those generated during the later radiation era.
Interestingly, for the same kination energy scale $\rho_{\rm kin}^{1/4}$, the fluctuations produced from the scenario with kination domination are enhanced by a factor of $\exp(3N_{\rm KD}/2)$, while the fluctuations in the case of subdominant fast-rolling field are suppressed by a factor $\Omega_{\phi}(\eta_{\rm kin})$.

\subsection{Application to the rotating axion}

For a nearly-quadratic potential Peccei-Quinn \eqref{quadraticpotential}, the rotating scalar field can lead to a kination era following a matter era, inside the radiation era,  \cite{Co:2019wyp,Co:2020jtv,Co:2021lkc,Gouttenoire:2021wzu,Gouttenoire:2021jhk}. 
This model was extensively studied in \cite{Gouttenoire:2021jhk}.
The energy density of the zero mode is obtained by scaling it as $a^{-6}$ from the start of kination era and switching to $a^{-3}$ when $\overline{\rho_{\phi}} \simeq m_a^2 f_a^2$ at $a_{\rm osc}$, similarly to the derivation of Eq.~\eqref{eq:rho_zero_end_final}. Its energy density today reads,
\begin{equation}
    \label{eq:Scalar_domination_rho_zero_mode_simple}
     \Omega_{\rm zero}^0 \simeq  \frac{1}{\sqrt{3}}\Omega_{\rm rad}^{0} \left(\frac{a_0 m_a}{\mathcal{H}_{\rm kin}}\right) \exp(2 N_{\rm KD}) \left(\frac{f_a}{M_{\rm Pl}}\right),
\end{equation}
The effectiveness of the curvature-induced axion production mechanism is described by the ratio,
\begin{equation}
    \frac{\Omega_{\rm fluct}^0}{\Omega_{\rm zero}^{0}} \simeq \sqrt{3}\left(\frac{M_{\rm pl}}{f_a}\right) \mathcal{P}_{\mathcal{R}}(k_{\rm kin}) \left( \frac{8}{\pi} N_{\rm KD} + \alpha^2\right),
\end{equation}
where we see that the fluctuation abundance is boosted slightly by $N_{\rm KD}$, instead of the suppression factor $\Omega_{\phi}^{1/2}(\eta_{\rm kin})$ in Eq.~\eqref{eq:yield_ratio} of the case of subdominant scalar-field.
Plugging in numerical values and $N_{\rm KD} \sim \mathcal{O}(1)$, we get
\begin{align}
    \frac{\Omega_{\rm fluct}^0}{\Omega_{\rm zero}^{0}} \simeq 2.26\, N_{\rm KD} \left(\frac{10^{10}\,{\rm GeV}}{f_a}\right) \left(\frac{\mathcal{P}_{\mathcal{R}}(k_{\rm kin})}{2.1 \times 10^{-9}}\right).
    \label{eq:ratio_fluct_zero_dominant_case}
\end{align}
Eq.~\eqref{eq:ratio_fluct_zero_dominant_case} suggests, like Eq.~\eqref{eq:yield_ratio}, that the curvature-induced axion DM becomes more prominent in the regime $f_a \lesssim 10^{10} ~ \textrm{GeV}$ for the scale-invariant $\mathcal{P}_{\mathcal{R}}(k)$.
We show the parameter space in Fig.~\ref{fig:kination_domination} where DM comes from fluctuations and the zero mode. Moreover, in this region of parameter space, the rotating axion leads to a period of kination era that imprints a detectable peak signature in the inflationary gravitational-wave background (see section 3.1 of \cite{Gouttenoire:2021jhk}).

\bibliographystyle{apsrev4-1}
\bibliography{roulette.bib}

\end{document}